\documentclass[%
 aip,
 sd,%
 amsmath,amssymb,
reprint,%
]{revtex4-1}

\usepackage{graphicx}% Include figure files
\usepackage{dcolumn}% Align table columns on decimal point
\usepackage{bm}% bold math
\usepackage{multirow}
\usepackage{color}

\begin{document}

\preprint{AIP/123-QED}

\title[]{Circular and linear magnetic
birefringences in xenon at $\lambda = 1064$\,nm}

\author{Agathe Cad\`ene}
%\email{xxx}
\affiliation{Laboratoire National des Champs Magn\'etiques
Intenses, (UPR 3228, CNRS-UPS-UJF-INSA), 143 avenue de Rangueil,
31400 Toulouse, France}

\author{Mathilde Fouch\'e}
%\email{xxx}
\affiliation{Laboratoire National des Champs Magn\'etiques
Intenses, (UPR 3228, CNRS-UPS-UJF-INSA), 143 avenue de Rangueil,
31400 Toulouse, France}

\author{Alice Riv\`ere}
\affiliation{Laboratoire National des Champs Magn\'etiques
Intenses, (UPR 3228, CNRS-UPS-UJF-INSA), 143 avenue de Rangueil,
31400 Toulouse, France}

\author{R\'emy
Battesti}
%\email{xxx}
\affiliation{Laboratoire National des Champs Magn\'etiques
Intenses, (UPR 3228, CNRS-UPS-UJF-INSA), 143 avenue de Rangueil,
31400 Toulouse, France}

\author{Sonia Coriani}
\affiliation{Dipartimento di Scienze Chimiche e Farmaceutiche,
Universit{\`a} degli Studi di Trieste, via Giorgieri 1, 34127
Trieste, Italy } \affiliation{Aarhus Institute of Advanced
Studies, Aarhus University, DK-8000 Aarhus C, Denmark}

\author{Antonio
Rizzo} \affiliation{Istituto per i Processi Chimico-Fisici del
Consiglio Nazionale delle Ricerche (IPCF-CNR), Area della Ricerca,
via G. Moruzzi 1, I-56124 Pisa, Italy}

\author{Carlo
Rizzo} \email{carlo.rizzo@lncmi.cnrs.fr} \affiliation{Laboratoire
National des Champs Magn\'etiques Intenses, (UPR 3228,
CNRS-UPS-UJF-INSA), 143 avenue de Rangueil, 31400 Toulouse,
France}

\date{\today}% It is always \today, today,

\begin{abstract}
The circular and linear magnetic birefringences corresponding to
the Faraday and the Cotton-Mouton effects, respectively, have been
measured in xenon at $\lambda = 1064$\,nm. The experimental setup
is based on time dependent magnetic fields and a high finesse
Fabry-P\'erot cavity. Our value of the Faraday effect is the first
measurement at this wavelength. It is compared to theoretical
predictions. Our uncertainty of a few percent yields an agreement
at better than 1$\sigma$ with the computational estimate when
relativistic effects are taken into account. Concerning the
Cotton-Mouton effect, our measurement, the second ever published
at $\lambda = 1064$\,nm, agrees at better than $1\sigma$ with
theoretical predictions. We also compare our error budget with
those established for other experimental published values.
\end{abstract}

%Use showkeys class option if keyword
\maketitle

%%%%%%%%%%%%%%%%%%%%%%%%%%%%%%%%%%%%%%%%%%%%%%%%%
\section{Introduction}
%%%%%%%%%%%%%%%%%%%%%%%%%%%%%%%%%%%%%%%%%%%%%%%%%
%\paragraph{This is the next level heading.~~} For this level please use \texttt{\textbackslash paragraph}. These headings should also end in a full point.

Magnetic birefringence corresponds to an anisotropy of the
(generally complex) refractive index induced in a medium by a
magnetic field.~\cite{barronbook,Rizzo_2005} A circular
birefringence arises when the magnetic field changes the angular
velocity of the two eigen modes of polarization in which a
linearly polarized beam is split, without deforming them. The net
result is a rotation of the plane of linear polarization, a
phenomenon seen also in absence of external fields in chiral
samples (natural optical rotation). When the  presence of the
external magnetic field yields a different phase of two
perpendicular components of the linear polarization vector, the
net result is the appearance of an ellipticity, and we are
observing an example of linear birefringence.

Two well known examples of magnetic birefringences are the Faraday
and the Cotton-Mouton effects. The former corresponds to a
circular birefringence induced by a longitudinal magnetic field
$B_\|$ (aligned parallel to the direction of propagation of
light). After going through the birefringent medium, the real part
of the index of refraction for left circularly polarized light
$n_-$ is different from that for right circularly polarized light
$n_+$. The difference $\Delta n_\mathrm{F} = n_- - n_+$ is
proportional to $B_\|$
\begin{equation}
\Delta n_\mathrm{F} = k_\mathrm{F}B_\|,\label{Eq:kF}
\end{equation}
$k_\mathrm{F}$ being the circular magnetic birefringence per unit
magnetic field intensity. For historical reason, the Faraday
effect is usually given in terms of the Verdet constant
\begin{eqnarray}
V = \frac{\pi k_\mathrm{F}}{\lambda},\label{Eq:k_F}
\end{eqnarray}
where $\lambda$ is the light wavelength. On the other hand, the
Cotton-Mouton effect corresponds to a linear magnetic
birefringence induced by a transverse magnetic field $B_\bot$. The
field induces a difference between the real parts of the
refraction index for light polarized parallel with respect to that
polarized perpendicular to the magnetic field. The difference
$\Delta n_\mathrm{CM} = n_\| - n_\bot$ is proportional to the
square of the magnetic field
\begin{equation}
\Delta n_\mathrm{CM} = k_\mathrm{CM}B_\bot^2,
\end{equation}
with $k_\mathrm{CM}$ the linear magnetic birefringence per square
unit magnetic field intensity.

For the Cotton-Mouton effect, $k_\mathrm{CM}$ has two
contributions, the first one due to the distortion of the
electronic structure while the second one corresponds to a partial
orientation of the molecules. When working in the conditions of
constant volume, the orientational contribution is proportional to
the inverse of the temperature $T$, and it usually dominates,
often hiding the first temperature independent contribution. For
axial molecules, for examples, $k_\mathrm{CM}$ is given by the
expression\cite{Rizzo_Rizzo}
\begin{equation}
\label{CME_birefringence} k_\mathrm{CM} = \frac{\pi
N_\mathrm{A}}{V_\mathrm{m}4\pi\epsilon_0}\Big(\Delta\eta +
\frac{2}{15 k_\mathrm{B}T}\Delta\alpha \Delta\chi \Big).
\end{equation}
Above $N_\mathrm{A}$ is the Avogadro constant, $V_\mathrm{m}$ the
molar volume, $k_\mathrm{B}$ the Boltzmann constant, $\epsilon_0$
the electric constant, $\Delta \eta$ the frequency dependent
hypermagnetizability anisotropy, $\Delta\alpha$ the optical
electric dipole polarizability anisotropy, and $\Delta\chi$ the
magnetic susceptibility anisotropy. For spherical molecules or for
atoms, such as xenon, however, the temperature dependent
contribution vanishes. Measurements on noble gases, for example,
allow to focus on the hypermagnetizability anisotropy $\Delta
\eta$ term. On the other hand, since the Langevin-type
orientational term vanishes, the magnetic birefringence is much
lower than the one observed in non spherical molecules. From an
experimental point of view, measurements on such gases require a
very sensitive apparatus, with a $\Delta n_\mathrm{CM}$ of the
order of $10^{-16}$ for helium and $10^{-14}$ for xenon at one
atmosphere and with a magnetic field of one Tesla. In comparison,
$\Delta n_\mathrm{F}$ is typically $10^{5}$ bigger.

The computational determination of the Verdet constant and of the
Cotton-Mouton effect requires the far-from-trivial calculation of
higher-order response functions,~\cite{Rizzo_Rizzo,Rizzo_2005} and
it has often served as test bed for the validation of new
electronic structure methods. For atoms, in order to obtain
accurate results one must properly account for the appropriate
description of one-electron (basis set), N-electron (correlation)
and relativistic effects. As far as correlation is concerned,
coupled cluster (CC) methods are nowadays among the most accurate
tools in electronic structure
theory.~\cite{Helgaker_2012,Christiansen_1998} Both birefringences
treated here, and in particular the Cotton-Mouton effect, require
a good description of the outer valence space of the system at
hand, and therefore the presence of diffuse functions in the
one-electron basis set is
mandatory.~\cite{Helgaker_2012,Rizzo_2005} Whereas for light atoms
relativistic corrections are minor, their importance increases and
they become significant for heavier atoms. For example, Ekstr\"om
\textit{et al}\cite{ekstrom_2005} have calculated that for helium
the relativistic effects add $-$0.03$\%$ to the non-relativistic
Verdet value. For xenon, the heaviest non-radioactive noble atom,
relativistic corrections add 3 to 4$\%$, depending on the chosen
wavelength. In this case, relativistic effects cannot be ignored
in accurate calculations.

In this article, we report both measurements and calculations of
Faraday and Cotton-Mouton effects at $\lambda = 1064$\,nm. We
perform the first measurement of the Faraday effect of xenon at
this wavelength, and our estimate bears an uncertainty of a few
percent. Concerning the Cotton-Mouton effect, our measurement, the
second ever published at $\lambda = 1064$\,nm, agrees at better
than $1\sigma$ with theoretical predictions and we also compare
our error budget with those established for other experimental
published values. Our theoretical predictions, that can be
considered of state-of-the-art quality, were obtained at the
coupled cluster singles and doubles
(CCSD)~\cite{CCSD:Bartlett,idCCSD:Koch2,CCSD:QR} and coupled
cluster singles, doubles and approximate triples
(CC3)~\cite{CC3,CC3:rsp,CC3:QR,CC3:Filip} levels of theory, and
they include estimates of relativistic effects. For both effects,
our theoretical predictions are within $1\sigma$ of our
experimental data.

%%%%%%%%%%%%%%%%%%%%%%%%%%%%%%%%%%%%%%%%%%%%%%%%%
\section{Experimental setup}
%%%%%%%%%%%%%%%%%%%%%%%%%%%%%%%%%%%%%%%%%%%%%%%%%

\subsection{Principle of the measurement}

Experimentally, we determine the Faraday and the Cotton-Mouton
effects by measuring, respectively, the rotation induced by a
longitudinal magnetic field and the ellipticity induced by a
transverse magnetic field on an incident linear polarization. For
small angles, the induced rotation $\theta_\mathrm{F}$ depends on
the circular birefringence as follows
\begin{eqnarray} \label{Eq:ThetaF_intro}
\theta_\mathrm{F} = \pi \frac{L_B}{\lambda} \Delta n_\mathrm{F},
\end{eqnarray}
where $L_B$ is the length of the magnetic field region. The
induced ellipticity $\psi_\mathrm{CM}$ is related to the linear
birefringence by the formula:
\begin{eqnarray} \label{Eq:Psi_intro}
\psi_\mathrm{CM} = \pi \frac{L_B}{\lambda} \Delta n_\mathrm{CM}
\sin2\theta_\mathrm{P},
\end{eqnarray}
where $\theta_\mathrm{P}$ is the angle between the light
polarization and the magnetic field.

\subsection{General setup}

\begin{figure}[h]
\centering
%\begin{center}
\includegraphics[width=8cm]{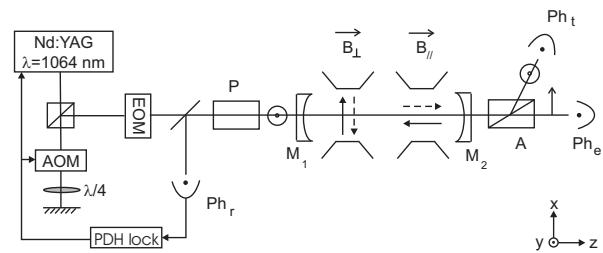}
\caption{\label{Fig:ExpSetup} Experimental setup. EOM =
electro-optic modulator; AOM = acousto-optic modulator; PDH =
Pound-Drever-Hall; Ph = photodiode; P = polarizer; A = analyzer.
See text for more details.}
%\end{center}
\end{figure}

The apparatus has already been described in detail
elsewhere.~\cite{Battesti_2008,Cadene_2013} Briefly, light comes
from a Nd:YAG laser at $\lambda = 1064$\,nm (see
Fig.\,\ref{Fig:ExpSetup}). It is linearly polarized by a first
polarizer P, before going through a transverse or a longitudinal
magnetic field. The polarization is then analyzed by a second
polarizer A, crossed at maximum extinction compared to P. The beam
polarized parallel to the incident beam, reflected by the
polarizer A as the ordinary ray, is collected by the photodiode
Ph$_\mathrm{t}$. Its power is denoted by $I_\mathrm{t}$. The beam
polarized perpendicular to the incident beam (power
$I_\mathrm{e}$), corresponding to the extraordinary ray that
passes through the polarizer A, is collected by the low noise and
high gain photodiode Ph$_\mathrm{e}$.

This setup has been designed to measure the linear magnetic
birefringence of vacuum \cite{Cadene_2014} and its sensitivity
allows to perform precise measurements on
gases.~\cite{Cadene_2013,Berceau_2012} All the optical components
from A to P are placed in an ultrahigh-vacuum chamber. To perform
birefringence measurement on gases, we fill the vacuum chamber
with a high-purity gas. For this particular measurement, we have
used a bottle of xenon with a global purity higher than
99.998\,$\%$.

\subsection{Fabry-P\'erot cavity}

Magnetic birefringence measurements on dilute gases are difficult,
especially at low pressure, because one has to detect very small
variations of light polarization. To increase the measured signal,
one needs high magnetic fields. One also needs an as large as
possible path length in the field $L_B$ (cf.
Eqs\,(\ref{Eq:ThetaF_intro}) and (\ref{Eq:Psi_intro})). To this
end, optical cavities are used to trap light in the magnetic field
region and therefore enhance the signal to be measured.

As shown in Fig.\,\ref{Fig:ExpSetup}, the cavity is formed by two
mirrors M$_1$ and M$_2$, placed at both sides of the magnetic
field region. The laser frequency is locked to the cavity
resonance frequency, using the Pound-Drever-Hall
technique.\cite{PDH_microwave} The electro-optic modulator
generates 10\,MHz sidebands and the signal reflected by the cavity
is detected by the photodiode Ph$_\mathrm{r}$. The laser frequency
is adjusted with the acousto-optic modulator, the piezoelectric
and the Peltier elements of the laser.

This cavity increases the distance traveled by light in the
magnetic field by a factor $2F/\pi$, where $F$ is the cavity
finesse. Therefore, the rotation induced by the longitudinal
magnetic field becomes
\begin{eqnarray}
\Theta_{\mathrm{F}}(t) =
\frac{2F}{\pi}\theta_{\mathrm{F}}(t),\label{Eq:Theta_theta}
\end{eqnarray}
with $\theta_\mathrm{F}$ the rotation acquired without any cavity.
In the same way, the ellipticity induced by the transverse
magnetic field becomes
\begin{eqnarray}
\label{Eq:Psi_psi} \Psi_\mathrm{CM}(t) =
\frac{2F}{\pi}\psi_\mathrm{CM}(t),
\end{eqnarray}
with $\psi_\mathrm{CM}$ denoting the ellipticity acquired without
any cavity. The cavity finesse is inferred from the measurement of
the photon lifetime $\tau$ inside the cavity\cite{Berceau_2010}
\begin{eqnarray}
F = 2\pi\Delta^\mathrm{FSR}\tau,
\end{eqnarray}
with $\Delta^\mathrm{FSR}$ the cavity free spectral range. For the
Faraday effect, the cavity finesse was about $F = 475\,000$. For
the Cotton-Mouton effect, two sets of mirrors were used with a
respective finesse of about 400\,000 and 480\,000.

\subsection{Raw signals}

We measure the circular and the linear magnetic birefringence by
measuring the ratio $I_\mathrm{e}/I_\mathrm{t}$
\begin{eqnarray}\label{Eq:Ie_It}
\frac{I_{\mathrm{e}}(t)}{I_{\mathrm{t,f}}(t)}=\sigma^2 + [\Gamma +
\Psi_\mathrm{CM}(t)]^2 + [\epsilon + \Theta_{\mathrm{F}}(t)]^2.
\end{eqnarray}
As said previously, $I_\mathrm{e}$ ($I_\mathrm{t}$) corresponds to
the power of light polarized perpendicular (parallel) to the
incident beam. The subscript f indicates that we need to take into
account the cavity filtering, as explained in details in previous
papers.\cite{Berceau_2010,Cadene_2013} The term $\sigma^2$
corresponds to the extinction ratio of polarizers P and A,
$\Gamma$ is the total static ellipticity due to the cavity mirrors
and $\epsilon$ is the static angle between the major axis of the
elliptical polarization and the incident polarization. The
extinction ratio and the static birefringence are measured before
each magnetic pulse. The static angle $\epsilon$ can be estimated
but its value is not needed for the analysis.

%%%%%%%%%%%%%%%%%%%%%%%%%%%%%%%%%%%%%%%%%%%%%%%%%
\section{Circular magnetic birefringence}
%%%%%%%%%%%%%%%%%%%%%%%%%%%%%%%%%%%%%%%%%%%%%%%%%

\subsection{Magnetic field}

The magnetic field is generated by a solenoid previously used for
Faraday effect measurement in helium.~\cite{Cadene_2013} Its
characteristics have already been explained in
details.~\cite{Cadene_2013} Here we just briefly recall its main
features. It generates a longitudinal magnetic field with an
equivalent length $L_B = (0.308 \pm 0.006)$\,m at 1$\sigma$. This
magnetic field is modulated at the frequency $\nu = 18$\,Hz: $B_\|
= B_{\|,0}\sin(2\pi \nu t + \phi)$. The rotation of the
polarization due to the Faraday effect is thus given by
\begin{eqnarray}
\Theta_{\mathrm{F}} &=& \Theta_0\sin(2\pi \nu t +\phi),\\
\mathrm{with}\; \Theta_0 &=& \frac{2F}{\pi} V B_{\|,0} L_B.
\label{Eq:Theta_0}
\end{eqnarray}

\subsection{Data analysis}

Expanding Eq.\,(\ref{Eq:Ie_It}), the raw signal becomes
\begin{equation}
\frac{I_{\mathrm{e}}(t)}{I_\mathrm{t,f}(t)} = \sigma^2 + \Gamma^2
+ \epsilon^2 +2\epsilon\Theta_\mathrm{F}(t) +
\Theta_\mathrm{F}^2(t).\label{Eq:Iext_It_general_Faraday}
\end{equation}
This gives three main frequency components: a DC signal, a signal
at the frequency $\nu$, and a signal at the double frequency
$2\nu$. To measure the Verdet constant, we use the amplitude of
the signal at $2\nu$~\cite{Cadene_2013}
\begin{eqnarray}
A_\mathrm{2\nu} &=& \frac{\Theta_0^2}
{2\sqrt{1+\big(\frac{2\nu}{\nu_\mathrm{c}}\big)^2}},
\label{Eq:Faraday_2f}
\end{eqnarray}
where $\nu_\mathrm{c}=1/4\pi\tau$ is the cavity cutoff frequency,
introduced to take into account the cavity
filtering.~\cite{Berceau_2010} $A_{2\nu}$ is measured for
different magnetic field amplitudes, from 0 to about $50\times
10^{-3}$\,T. The whole is fitted by $K_V B_{\|,0}^2$. The Verdet
constant finally depends on the measured experimental parameters
as follows
\begin{equation}
V(T,P) = \sqrt{\frac{K_V}{2}}
\frac{\Big[1+{(8\pi\tau\nu)}^{2}\Big]^{1/4}}
{2\tau\Delta^{\mathrm{FSR}}L_B},\label{Eq:Verdet}
\end{equation}
where $T$ and $P$ are respectively the temperature and pressure of
the gas.

\subsection{Measurement and error budget}

The A- and B-type uncertainties associated to the measurement of
$V$ are detailed in
Tab.\,\ref{Tab:TableBudget}.~\cite{Berceau_2012,Cadene_2013} They
are given at $1\sigma$ (coverage factor $k = 1$). The A-type
uncertainty is dominated by the photon lifetime uncertainty. The
main contributions of the B-type uncertainty comes from the
uncertainty of the magnetic length and of the fit constant $K_V$
which includes the B-type uncertainty of the magnetic field and of
the photodiodes conversion factor~\cite{Berceau_2012}.

\begin{table}
\caption{\label{Tab:TableBudget} Parameters and their respective
relative A- and B-type uncertainties at 1$\sigma$ that have to be
measured to infer the value of the Verdet constant $V$. Typical
values are given at $P = 5\times 10^{-3}$\,atm.}
\begin{center}
%\begin{tabular*}{1\textwidth}{@{\extracolsep{\fill}}{c c c c c c}
\begin{tabular*}{0.47\textwidth}{p{1.9cm} p{1.9cm} p{2cm}  p{2cm}}
  \hline
  \hline
  \centering Parameter & Typical &  Relative A-type & Relative B-type\\
  & value & uncertainty & uncertainty\\
 \hline
& & &\\
 $\tau$ [ms] &    1.14 &  \raggedright $2.0\times 10^{-2}$   & \\
$K_V$ [rad T$^{-1}$] &      1.07      &  $3\times 10^{-3}$ & $3.2\times 10^{-2}$\\
$\Delta^{\mathrm{FSR}}$ [MHz] & 65.996 &  &  $3\times 10^{-4}$   \\
$L_B$ [m] &   0.308     &   &  $1.9\times 10^{-2}$   \\
 \hline
& & &\\
$V\times 10^{5}$ & 1.66  & $1.8\times 10^{-2}$ &  $2.5\times 10^{-2}$ \\
$[\mathrm{rad}\mathrm{T}^{-1}\mathrm{m}^{-1}]$ & & & \\
 \hline
 \hline
\end{tabular*}
\end{center}
\end{table}

We have measured the Verdet constant in xenon at $T=(294\pm1)$\,K
and for 5 pressures from $1.01\times 10^{-3}$ to $5.01 \times
10^{-3}$\,atm. In this range of pressure, xenon can be considered
as an ideal gas and the Verdet constant is thus proportional to
the pressure. Data are fitted by a linear equation:

\begin{equation}\label{Eq:V_fit_P}
V(T,P) = V^{\mathrm{n}} P,
\end{equation}
giving a normalized Verdet constant ($P = 1$\,atm) at $\lambda =
1064$\,nm and $T = (294 \pm 1)$\,K
\begin{equation}\label{Eq:V_293K}
V^{\mathrm{n}} = (3.31 \pm 0.09)\times 10^{-3}\;
\mathrm{atm}^{-1}\mathrm{rad\ T}^{-1}\mathrm{m}^{-1}.
\end{equation}
The uncertainty is given at $1\sigma$ and is detailed in
Tab.\,\ref{Tab:TableBudget2}. With a scale law on the gas density,
this corresponds to a normalized Verdet constant at $T =
273.15$\,K of
\begin{equation}\label{Eq:V_293K_2}
V^{\mathrm{N}} = (3.56 \pm 0.10)\times 10^{-3}\;
\mathrm{atm}^{-1}\mathrm{rad\ T}^{-1}\mathrm{m}^{-1}.
\end{equation}
Using Eq.\,(\ref{Eq:k_F}), we can also give the normalized Faraday
constant at $T = 273.15$\,K
\begin{equation}
k_\mathrm{F}^{\mathrm{N}} = (1.21 \pm 0.03)\times 10^{-9}\;
\mathrm{atm}^{-1}\mathrm{T}^{-1}.
\end{equation}

\begin{table}
\caption{\label{Tab:TableBudget2} Parameters and their respective
relative A- and B-type uncertainties at 1$\sigma$ that have to be
measured to infer the value of the normalized Verdet constant
$V^\mathrm{n}$. The uncertainty given by the linear fit takes into
account the A-type uncertainty of $V$.}
\begin{center}
%\begin{tabular*}{1\textwidth}{@{\extracolsep{\fill}}{c c c c c c}
\begin{tabular*}{0.47\textwidth}{p{1.9cm} p{1.9cm} p{2cm}  p{2cm}}
  \hline
  \hline
  \centering Parameter & Typical &  Relative A-type & Relative B-type\\
  & value & uncertainty & uncertainty\\
 \hline
& & &\\
$V \times 10^{5}$ & 1.66  & $1.8\times 10^{-2}$ &  $2.5\times 10^{-2}$ \\
$[\mathrm{rad}\mathrm{T}^{-1}\mathrm{m}^{-1}]$ & & & \\
$P\times 10^{3}$ &      5      &  & $2\times 10^{-3}$\\
$[$atm$]$ & & &\\
linear fit  & $3.31$ & $1.5 \times 10^{-2}$ &    \\
$\times 10^{3}$ & & & \\
$[\mathrm{atm}^{-1}\mathrm{rad}$ & & & \\
$\mathrm{T}^{-1}\mathrm{m}^{-1}]$ & & & \\
 \hline
& & &\\
$V^\mathrm{n} \times 10^{3}$ & 3.31  & $1.5\times 10^{-2}$ &  $2.5\times 10^{-2}$ \\
$[\mathrm{atm}^{-1}\mathrm{rad}$ & & & \\
$\mathrm{T}^{-1}\mathrm{m}^{-1}]$ & & & \\
 \hline
 \hline
\end{tabular*}
\end{center}
\end{table}

%%%%%%%%%%%%%%%%%%%%%%%%%%%%%%%%%%%%%%%%%%%%%%%%%
\section{Linear magnetic birefringence}
%%%%%%%%%%%%%%%%%%%%%%%%%%%%%%%%%%%%%%%%%%%%%%%%%

\subsection{Magnetic field}

The transverse magnetic field $B_{\perp}$ is generated by an
X-Coil, specially designed by the High Magnetic Field National
Laboratory (LNCMI-Toulouse, France) for the measurement of the
vacuum magnetic birefringence. This coil has been presented and
discussed in great details in several previous
papers.~\cite{Battesti_2008,Batut-2008} Very briefly, the magnet
delivers a pulsed magnetic field over an equivalent length $L_B$
of 0.137\,m. The total duration of the pulse is about 10\,ms with
a maximum reached within 2\,ms. For the present measurements, a
maximum magnetic field of 3\,T has been used. Finally, the
high-voltage connections can be remotely switched to reverse the
direction of the field. Thus we can set $\textbf{B}$$_{\perp}$
parallel or antiparallel to the $x$ direction, as shown in
Fig.\,\ref{Fig:ExpSetup}.

\subsection{Data analysis}

The data analysis follows the one described for the Cotton-Mouton
effect measurement in helium.~\cite{Cadene_2013} We will however
detail the main steps, since a slightly different method was used
in the present case.

To extract the ellipticity $\Psi_\mathrm{CM}(t)$ from
Eq.\,(\ref{Eq:Ie_It}), we calculate the following $Y(t)$ function
\begin{eqnarray}\label{Eq:Y(t)}
Y(t) &=& \frac{\frac{I_{\mathrm{e}}(t)}{I_{\mathrm{t,f}}(t)} - I_\mathrm{DC}}{2|\Gamma|} \nonumber \\
&=& \gamma \Psi_\mathrm{CM}(t) +
\frac{\Psi_\mathrm{CM}^2(t)}{2|\Gamma|}+\gamma
\frac{|\epsilon|\Theta_{\mathrm{F}}(t)}{2|\Gamma|} +
\frac{\Theta_{\mathrm{F}}^2(t)}{2|\Gamma|}, \nonumber \\
\label{Eq:Y(t)_2}
\end{eqnarray}
where $\gamma$ stands for the sign of $\Gamma$. $I_\mathrm{DC}$ is
the static signal measured just before the application of the
magnetic field. The absolute value of the static ellipticity
$|\Gamma|$ is also measured before each pulse.

Two parameters are adjustable in the experiment: the sign $\gamma$
of the static ellipticity $\Gamma$ and the direction of the
transverse magnetic field. We acquire signals for both signs of
$\Gamma$ and both directions of $\textbf{B}$$_{\perp}$: parallel
to $x$ is denoted as $>0$ and antiparallel is denoted as $<0$.
This gives four data series: ($\Gamma>0$, $B_{\perp}>0$),
($\Gamma>0$, $B_{\perp}<0$), ($\Gamma<0$, $B_{\perp}<0$) and
($\Gamma<0$, $B_{\perp}>0$).

For each series, signals calculated with Eq.\,(\ref{Eq:Y(t)}) are
averaged and denoted as $Y_{>>}$, $Y_{><}$, $Y_{<<}$ and $Y_{<>}$.
The first subscript corresponds to $\Gamma
>0$ or $<0$ while the second one corresponds to
$\textbf{B}$$_{\perp}$ parallel or antiparallel to $x$.. This
average function can be written in a more general form than the
one of Eq.\,(\ref{Eq:Y(t)_2}). It is the sum of different effects
with different symmetries, denoted as $s$
\begin{eqnarray}
\nonumber Y_{>>} &=& +\Psi +
\frac{1}{2}\Big<\frac{1}{\Gamma_{>>}}\Big>s_{++} +
\Big<\frac{1}{\Gamma_{>>}}\Big>s_{--} +
\frac{1}{2}\Big<\frac{1}{\Gamma_{>>}}\Big>s_{+-},\\
\nonumber Y_{><} &=& +\Psi +
\frac{1}{2}\Big<\frac{1}{\Gamma_{><}}\Big>s_{++} +
\Big<\frac{1}{\Gamma_{><}}\Big>s_{--} +
\frac{1}{2}\Big<\frac{1}{\Gamma_{><}}\Big>s_{+-},\\
\nonumber Y_{<<} &=& -\Psi +
\frac{1}{2}\Big<\frac{1}{\Gamma_{<<}}\Big>s_{++} +
\Big<\frac{1}{\Gamma_{<<}}\Big>s_{--} +
\frac{1}{2}\Big<\frac{1}{\Gamma_{<<}}\Big>s_{+-},\\
\nonumber Y_{<>} &=& -\Psi +
\frac{1}{2}\Big<\frac{1}{\Gamma_{<>}}\Big>s_{++} +
\Big<\frac{1}{\Gamma_{<>}}\Big>s_{--} +
\frac{1}{2}\Big<\frac{1}{\Gamma_{<>}}\Big>s_{+-}.\\
\label{Eq:Y}
\end{eqnarray}
The first subscript in $s$ corresponds to the symmetry with
respect to the sign of $\Gamma$ and the second one to the symmetry
with respect to the direction of $\textbf{B}$$_{\perp}$. The
subscript $+$ indicates an even parity while the subscript $-$
indicates odd parity. The ratio $<1/\Gamma>$ is the average of
$1/|\Gamma|$ measured during corresponding series. The terms
$\Psi_\mathrm{CM}^2$ and $\Theta_\mathrm{F}^2$ are included in
$s_{++}$, $\gamma|\epsilon|\Theta_\mathrm{F}$ are included in
$s_{--}$, and $s_{+-}$ corresponds to a spurious signal with an
odd parity towards the direction of $\textbf{B}$$_{\perp}$ and an
even parity with respect to the sign of $\Gamma$. The ellipticity
$\gamma\Psi_\mathrm{CM}$ corresponds to $s_{-+}$.

From this set of four equations with four unknown quantities
($\Psi_\mathrm{CM}$, $s_{++}$, $s_{--}$ and $s_{+-}$), we extract
$\Psi_\mathrm{CM}(t)$, which is fitted by $\alpha
B_{\perp,\mathrm{f}}^2$. The cavity filtering should again be
taken into account, as indicated by the subscript
$\mathrm{f}$.\cite{Berceau_2010,Cadene_2013} The Cotton-Mouton
constant $k_\mathrm{CM}$ finally depends on the measured
experimental parameters as follows:
\begin{eqnarray}\label{Eq:kcm}
k_{\mathrm{CM}}(T,P) = \frac{\alpha}{4\pi \tau
\Delta^\mathrm{FSR}}\frac{\lambda}{L_B}\frac{1}{\sin2\theta_\mathrm{P}}.
\end{eqnarray}

\subsection{Measurement and error budget}

The A- and B-type uncertainties associated to the measurement of
$k_\mathrm{CM}$ are detailed in Tab.\,\ref{Tab:Err_CM} and are
given at $1\sigma$. The B-type uncertainties have been evaluated
previously and detailed in Ref.\,\citenum{Berceau_2012}. They
essentially come from the length of the magnetic field
$L_\mathrm{B}$ and the fit constant $\alpha$.

\begin{table}
\caption{\label{Tab:Err_CM} Parameters that have to be measured to
infer the value of the Cotton-Mouton constant $k_{\mathrm{CM}}$
and their respective relative A- and B-type uncertainties at
1$\sigma$. Typical values are given at $P = 8\times
10^{-3}$\,atm.}
\begin{center}
%\begin{tabular*}{1\textwidth}{@{\extracolsep{\fill}}{c c c c c c}
\begin{tabular*}{0.47\textwidth}{p{1.9cm} p{1.9cm} p{2cm}  p{2cm}}
  \hline
  \hline
  \centering Parameter & Typical &  Relative A-type & Relative B-type\\
  & value & uncertainty & uncertainty\\
 \hline
& & &\\
 $\tau$ [ms] &    1.14 &  \raggedright $2.0\times 10^{-2}$   & \\
$\alpha \times 10^{5}$ $[$T$^{-2}]$ &      $2.82$      &  $2.8 \times 10^{-4}$ & $2.2 \times 10^{-2}$\\
$\Delta^{\mathrm{FSR}}$ [MHz] & 65.996 &  &  $3\times 10^{-4}$   \\
$L_{B}$ [m]    &  0.137   &    &  $2.2 \times 10^{-2}$   \\
$\lambda$ [nm]   &  1064.0   &    &  $<5 \times 10^{-4}$   \\
$\sin 2\theta_\mathrm{P}$   &   1.0000 & &  $9 \times 10^{-4}$   \\
 \hline
& & &\\
$k_\mathrm{CM}$ & $2.31$  & $2.0\times 10^{-2}$ &  $3.1\times 10^{-2}$ \\
$\times 10^{16}$ [T$^{-2}$] & & &\\
 \hline
 \hline
\end{tabular*}
\end{center}
\end{table}

We have measured the Cotton-Mouton constant in xenon at $T =
(293\pm 1)$\,K and for nine pressures from $3\times 10^{-3}$ to
$8\times 10^{-3}$\,atm. The data as a function of the pressure are
fitted by a linear equation, and we obtain for the value of the
Cotton-Mouton constant at $P = 1$\,atm
\begin{eqnarray}
k_{\mathrm{CM}}^\mathrm{n} = (2.41 \pm 0.37)\times
10^{-14}\,\mathrm{T^{-2}atm^{-1}}.
\end{eqnarray}
The uncertainty given at 1$\sigma$ is detailed in
Tab.\,\ref{Tab:Err_CM2}. The dominant uncertainty comes from the
linear fit of the Cotton-Mouton constant versus pressure (A-type).
The value of $k_{\mathrm{CM}}^\mathrm{n}$ normalized at 273.15\,K
is calculated with a scale law on the gas density
\begin{eqnarray}
k_{\mathrm{CM}}^\mathrm{N} = (2.59 \pm 0.40)\times
10^{-14}\,\mathrm{T^{-2}atm^{-1}}.
\end{eqnarray}

\begin{table}
\caption{\label{Tab:Err_CM2} Parameters and their respective
relative A- and B-type uncertainties at 1$\sigma$ that have to be
measured to infer the value of the normalized Cotton-Mouton
constant $k_{\mathrm{CM}}^\mathrm{n}$.}
\begin{center}
\begin{tabular*}{0.47\textwidth}{p{1.9cm} p{1.9cm} p{2cm}  p{2cm}}
  \hline
  \hline
  \centering Parameter & Typical &  Relative A-type & Relative B-type\\
  & value & uncertainty & uncertainty\\
 \hline
& & &\\
$k_\mathrm{CM}$ & $2.31$  & $2.0\times 10^{-2}$ &  $3.1\times 10^{-2}$ \\
$\times 10^{16}$ [T$^{-2}$] & & &\\
$P\times 10^{3}$ &      5      &  & $2\times 10^{-3}$\\
$[$atm$]$ & & &\\
linear fit  & 2.41 & $1.5\times 10^{-1}$ &    \\
$\times 10^{14}$ & & &\\
$[\mathrm{T}^{-2}\mathrm{atm}^{-1}]$ & & & \\
 \hline
& & &\\
$k_{\mathrm{CM}}^\mathrm{n}$ & 2.41  & $1.5\times 10^{-1}$ &  $3.1\times 10^{-2}$ \\
$\times 10^{14}$ & & &\\
$[\mathrm{T}^{-2}\mathrm{atm}^{-1}]$ & & & \\
 \hline
 \hline
\end{tabular*}
\end{center}
\end{table}

\section{Our calculations}
The Verdet constant and the Cotton-Mouton birefringence were
computed within Coupled Cluster response
theory,~\cite{Christiansen_1998,Helgaker_2012} at the
CCSD~\cite{CCSD:Bartlett,idCCSD:Koch2,CCSD:QR} and
CC3~\cite{CC3,CC3:rsp,CC3:QR,CC3:Filip} levels of approximation.
Specifically, the Verdet constant was obtained from the following
frequency-dependent quadratic response
function~\cite{Coriani_1997,Coriani_erratum,Coriani_2000,Helgaker_2012}
\begin{equation}
V(\omega) = C \omega
%\epsilon_{\alpha\beta\gamma}
\langle\langle \mu_x; \mu_y, L_{z}\rangle\rangle_{\omega,0},
\end{equation}
with $C=\frac{N e}{8 m_e \epsilon_0 c_0} = 0.912742\times 10^{-7}$
in atomic units, $N$ the number density
($N=\frac{P}{k_\mathrm{B}T}$ for ideal gases), $e$ the elementary
charge, $m_e$ the electron mass, $c_0$ the speed of light in
vacuo, $\omega/2\pi$ the frequency of the probing light, and
$\mu_{x,y}$ and $L_{z}$ are Cartesian components of the electric
dipole, and angular momentum operators, respectively. The
hypermagnetizability anisotropy $\Delta\eta$ entering the
Cotton-Mouton birefringence in Eq.~(\ref{CME_birefringence}) (the
only term contributing for atoms) is given by the combination of a
quadratic and a cubic response functions~\cite{Rizzo_Rizzo}
\begin{align}
\Delta\eta &= -\frac{1}{4} \langle\langle \mu_{x}; \mu_{x},
L_{z},L_{z} \rangle\rangle_{\omega,\omega,0} -\frac{1}{4}
\langle\langle \mu_{x}; \mu_{x}, \Theta_{xx}
\rangle\rangle_{\omega,0}\\\nonumber &\equiv \Delta\eta^p +
\Delta\eta^d.
\end{align}
with $\Theta_{xx}$ the Cartesian component of the traceless
quadrupole operator. At the CC3 level, calculations were performed
at three different wavelengths, namely 1064, 632.8 and 514.5 nm.
At the CCSD level, we computed the dispersion coefficients, as
done in our previous study,~\cite{Coriani_1999} i.e., for the
Verdet constant
\begin{align}
V(2n) & =  2n S(-2n-2); \\
V(\omega) & = C \sum_{n=1}^{\infty} \omega^{2n} V(2n);
\end{align}
whereas for the Cotton-Mouton constant
\begin{align}
\Delta\eta (2n) &= -\frac{1}{4} [(2n+1)(2n+2)S(-2n-4)+B(2n)];\\
\Delta\eta(\omega) &= \sum_{n=0}^{\infty} \omega^{2n}
\Delta\eta(2n).
\end{align}

Above, $S(k)$ is the Cauchy moment
\begin{equation}
S(k) = \sum_{m \ne 0} 2 \omega_{m0}^{k+1} \langle 0 \mid \mu_z
\mid m \rangle \langle m \mid \mu_z \mid 0 \rangle
\end{equation}
with $\hbar \omega_{m0}$ indicating the excitation energy from the
ground state $0$ to the excited state(s) $m$,  and $B(2n)$ is the
dispersion coefficient introduced when expanding, for frequencies
below the lowest excitation energy, the electric dipole--electric
dipole--electric quadrupole quadratic response function
$B_{x,x,xx}(-\omega;\omega,0) = \langle\langle \mu_{x}; \mu_{x},
\Theta_{xx} \rangle\rangle_{\omega,0}$ in a convergent power
series in the circular frequency $\omega$
\begin{equation}
B_{x,x,xx}(-\omega;\omega,0) = \sum_{n=0}^{\infty} \omega^{2n}
B(2n)
\end{equation}
For further details on how the above Cauchy moments and dispersion
coefficients of the given quadratic response function are computed
within coupled cluster response theory, the reader should refer to
Refs.~\onlinecite{Haettig_1997,Coriani_1999,Haettig_1998}.

Relativistic effects were approximately accounted for by employing
relativistic effective core potentials (ECPs),~\cite{Dolg_2000}
and specifically pseudo-potentials (PP). ``Small core'' effective
pseudo-potentials were used to describe the 28 inner electrons
(that is, the [Ar]3d$^{10}$  core), whereas the remaining 26
valence electrons were correlated as in standard non-relativistic
calculations. The basis sets used were constructed starting from
the singly augmented aug{\_}cc{\_}pvxz{\_}pp (x=t,q) sets of
Peterson {\em et al}.~\cite{Peterson_2003} Since single
augmentation is usually not sufficient to ensure converged
results, at least for the Cotton-Mouton birefringence. Additional
sets of diffuse functions were added by applying an even-tempered
generation formula commonly used for this purpose to the orbital
functions describing the valence electrons, while retaining the
pseudo-potential of the original set. The resulting sets are
labeled d-aug and t-aug, for double and triple augmentation,
respectively.

Where pseudo-potentials parametrically account for relativistic
effects on the innermost orbitals, other relativistic effects
(e.g. higher-order and picture change effects, spin-orbit
coupling) could play a significant
role.~\cite{Saue_2005,Saue_2011} When dealing with valence
properties like electric hyperpolarizabilities, the higher-order
relativistic effects and picture change effects (for the dipole
operator and also the electron-electron interaction) are expected
to be not so important. Also, spin-orbit coupling should be quite
weak. Both the Faraday and Cotton Mouton birefringences, however,
involve the magnetic dipole operator. In general relativistic
effects on magnetic properties can be more significant and more
difficult in terms of picture change (the operators look different
in relativistic and non-relativistic theory and this may require a
correction of the property operator that one uses as a
perturbation).~\cite{Saue_2005,Saue_2011}

Nonetheless, also given that the most stringent requirement in
terms of basis set convergence is the inclusion of diffuse
functions as in the case of the electric hyperpolarisability,
%also in view of where we need to converge the basis set,
it is reasonable to assume that both properties are essentially
valence properties, for which picture change effects are typically
small, and we reckon therefore that  the use of (PP)ECPs can be
considered accurate enough.

The results obtained in the x=q basis sets are summarized in
Tab.\,\ref{Tab:CCSD} and Tab.\,\ref{Tab:CC3}, for CCSD and CC3,
respectively.

All calculations were performed with the Dalton
code.~\cite{dalton}

\begin{table*}
\small \caption{\label{Tab:CCSD} Dispersion coefficients of the
Verdet and Cotton-Mouton response functions at the CCSD level of
theory (atomic units).}
\begin{center}
\begin{tabular*}{1\textwidth}{@{\extracolsep{\fill}}lccccc}
%\begin{tabular}{lccccc}
\hline\hline
$n$& $B(2n)$&$S(-2n-4)$&$S(-2n-2)$&$V(2n)$ & $\Delta\eta(2n)$ \\
\hline
\multicolumn{6}{c}{aug\_cc\_pvqz\_pp}\\
\hline
0& $-$654.89471 &  126.50595  &                   &                    &       100.47070 \\
1& $-$8903.3825 &  763.59899  & 126.50595 & 253.01190  & $-$64.951345 \\
2& $-$92298.251 &  5369.0486  & 763.59899 & 3054.3960  & $-$17193.302 \\
3& $-$860869.9   &  41692.560  & 5369.0486 & 32214.292  & $-$368478.36 \\
4&                         &                     & 41692.560 & 333540.48  &                         \\
\hline
\multicolumn{6}{c}{d-aug\_cc\_pvqz\_pp}\\
\hline
0   & $-$739.15630  &  126.97174   &                   &                   &       121.30323 \\
1   & $-$9822.9127  &  774.87190   & 126.97174 &  253.94348&       131.11247 \\
2   & $-$106369.19  &  5553.2321   & 774.87190 &  3099.4876& $-$15056.943 \\
3   & $-$1074975.1  &  44095.369   & 5553.2321 &  33319.393& $-$348591.39 \\
4   &     &                   &  44095.369   & 352762.95 &\\
\hline
\multicolumn{6}{c}{t-aug\_cc\_pvqz\_pp}\\
\hline
0  &  $-$748.34187 &   126.91927   &                   &                    &       123.62583\\
1   & $-$9940.7218 &   774.47234   & 126.91927 & 253.83854  &       161.76343\\
2   & $-$107513.63 &   5551.3771   & 774.47234 & 3097.8894  & $-$14756.921  \\
3   & $-$1084127.2 &   44088.280   & 5551.3771 & 33308.263  & $-$346204.12\\
4   &                         &                       & 44088.280 & 352706.24  & \\
\hline \hline
\end{tabular*}
\end{center}
\end{table*}

%\clearpage
%\normalsize

\begin{table*}
\small \caption{\label{Tab:CC3} CC3 values of the response
function components (in atomic units) involved in the Verdet and
Cotton-Mouton birefringences. The Verdet constant
$V^{\mathrm{N}}(\omega)$ is given in atm$^{-1}$ rad.
T$^{-1}$m$^{-1}$ and the Cotton-Mouton constant
$k_{\rm{CM}}^{\mathrm{N}}$ is in T$^{-2}$.atm$^{-1}$ at 273.15 K.}
\begin{center}
\begin{tabular*}{1\textwidth}{@{\extracolsep{\fill}}lcccccc}
%\begin{tabular}{ccccccc}
\hline\hline $\lambda$[nm] & $\langle\langle \mu_x;
\mu_y,L_z\rangle\rangle_{\omega,0}$  &
$V^{\mathrm{N}}(\omega)\times 10^{3}$ & $\langle\langle \mu_{x};
\mu_{x}, \Theta_{xx} \rangle\rangle_{\omega,0}$ & $\langle\langle
\mu_{x}; \mu_{x}, L_{z},L_{z} \rangle\rangle_{\omega,\omega,0}
$ & $\Delta\eta $ & $k_{\rm{CM}}^{\mathrm{N}}\times 10^{14}$\\
\hline
\multicolumn{7}{c}{aug\_cc\_pvqz\_pp} \\
\hline
1064  & 11.1587 & 3.505  & $-$668.242 & 272.564 & 98.9195  & 2.239\\
632.8 & 19.5823 & 10.35   &$-$700.706 & 308.069 & 98.1593 & 2.222\\
514.5 & 24.9438 & 16.22   &$-$728.260 & 339.617 & 97.1607 & 2.200 \\
%514.5 & 24.9438 & 16.22   &$-$728.260 & 339.617 & $-$97.1607 & $-$2.1996 \\
\hline
\multicolumn{7}{c}{d-aug\_cc\_pvqz\_pp}\\
\hline
1064  & 11.2155& 3.522  &$-$755.936 &   274.099 &120.459   & 2.727\\
632.8 & 19.6927& 10.40  &$-$791.994&    310.285 &120.427   & 2.726\\
514.5 & 25.0963& 16.32  &$-$822.705&    342.514 &120.048   & 2.718\\
\hline
\multicolumn{7}{c}{t-aug\_cc\_pvqz\_pp}\\
\hline
1064  & 11.2127&3.521  &$-$765.680 &    274.031 &122.912   & 2.782\\
632.8 & 19.6878&10.40  &$-$802.186 &    310.210 &122.994   & 2.784\\
514.5 & 25.0901&16.31  &$-$833.274 &    342.435 &122.710   & 2.778\\
\hline\hline
\end{tabular*}
\end{center}
\end{table*}

\section{Results and discussion}

\subsection{Faraday effect}

\subsubsection{Experiments}

\begin{figure}
\begin{center}
\includegraphics[width=8cm]{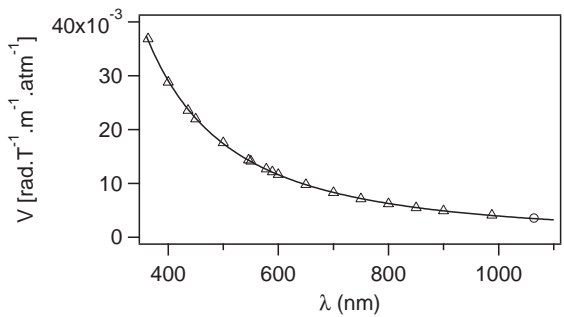}
\caption{\label{Fig:Comparaison_Exp_Faraday} $\vartriangle$:
Experimental values of xenon normalized Verdet constant at $T =
273.15$\,K reported by Ingersoll and
Liebenberg\cite{Ingersoll_1956} for wavelength from 363\,nm to
987.5\,nm. These values are fitted by the law
$A/\lambda^2+B/\lambda^4$ (solid line). $\circ$: Our experimental
value at $T = 273.15$\,K.}
\end{center}
\end{figure}

We can compare our value of the normalized Verdet constant to
other published values. The most extensive experimental
compilation of Verdet constants has been reported by Ingersoll and
Liebenberg in 1956, for several gases including
xenon\cite{Ingersoll_1956} for wavelengths ranging from 363.5 to
987.5\,nm, with a total uncertainty of about 1\,$\%$. These values
are plotted in Fig.\,\ref{Fig:Comparaison_Exp_Faraday}.

No datum has ever been reported for $\lambda = 1064$\,nm.
Nevertheless, we can extrapolate its value from the points of
Fig.\,\ref{Fig:Comparaison_Exp_Faraday}, by fitting the data with
a function of form $V = A/\lambda^2+B/\lambda^4$ (solid curve in
Fig.\,\ref{Fig:Comparaison_Exp_Faraday}).\cite{Rosenfeld_1929,Ingersoll_1956}
A supplementary systematic uncertainty should also be added, since
the authors measured the ratio between Faraday effects in xenon
and in distilled water, and rescaled their measurements with
accepted values for water.\cite{Rosenfeld_1929,Ingersoll_1956}.
Thus it does not correspond to absolute measurements of the
Faraday effect, contrary to ours.

At $\lambda = 1064$\,nm and $T = 273.15$\,K we obtain
$V^{\mathrm{N}} = (3.46 \pm 0.04)\times
10^{-3}$\,atm$^{-1}$rad.T$^{-1}$m$^{-1}$. The $1\sigma$
uncertainty includes the one given by the fit. This value is
compatible with our experimental value (Eq.\,(\ref{Eq:V_293K_2})),
represented as the open circle in
Fig.\,\ref{Fig:Comparaison_Exp_Faraday} and as the straight and
dashed lines in Fig.\,\ref{Fig:Comparaison_Faraday}.

\subsubsection{Theory}

\begin{figure}
\begin{center}
\includegraphics[width=8cm]{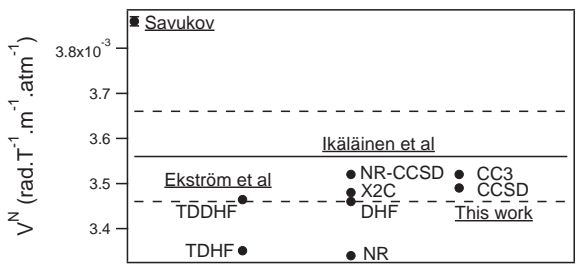}
\caption{\label{Fig:Comparaison_Faraday} Normalized Verdet
constant of xenon at $T = 273.15$\,K at $\lambda = 1064$\,m. Solid
line: our experimental mean value. Dashed lines: our experimental
value with 1$\sigma$ uncertainty. Points : theoretical predictions
(both ours and from the literature). See text and
Tab.\,\ref{Tab:ComparaisonFaraday} for the references.}
\end{center}
\end{figure}

We can also compare our experimental value with theoretical
predictions (both ours and from the literature), plotted in
Fig.\,\ref{Fig:Comparaison_Faraday} and summarized in
Tab.\,\ref{Tab:ComparaisonFaraday} at 1\,atm, 273.15\,K and with
the gas number density of an ideal gas. To convert from
theoretical results given in atomic units into the units used
experimentally, we exploited the relation:
\begin{equation}
V\ (\mathrm{atm}^{-1}\mathrm{rad.}\mathrm{T}^{-1}\mathrm{m}^{-1})
= V\ \mathrm{(a.u.)} \times 8.039617\times 10^4.
\end{equation}

Our experimental value is compatible within 1$\sigma$ with both
our ``best'' coupled cluster results (t-aug\_cc\_pvqz\_pp basis)
and the theoretical prediction of Ekstr\"om \textit{et
al},\cite{ekstrom_2005} within 2$\sigma$ with the estimate of
Ik\"al\"ainen \textit{et al},\cite{Ikalainen_2012}
%(\revSC{Dirac-Hartree Fock value}),
and within 3$\sigma$ with that of Savukov.\cite{Savukov_2012}

%{(\revSC{Configuration interaction?})}.
%I. M. Savukov\cite{Savukov_2012}.

\begin{table}
\caption{\label{Tab:ComparaisonFaraday} Experimental  and
theoretical values of the normalized Verdet constant at $T =
273.15$\,K, $\lambda = 1064$\,nm, with uncertainties at
1$\sigma$.}
\begin{center}
%\begin{tabular*}{1\textwidth}{@{\extracolsep{\fill}}{c c c c c c}
\begin{tabular*}{0.47\textwidth}{p{3cm} p{2cm}  p{5cm}}
  \hline
  \hline
  Ref. & $V^{\mathrm{N}}\times 10^{3}$  & Remarks\\
 & ($\mathrm{atm}^{-1}\mathrm{rad.}$
& \\
& $\mathrm{T}^{-1}\mathrm{m}^{-1}$) & \\
 \hline
 \textbf{Experiment} &  & \\
 & & \\
Ingersoll \textit{et al}\cite{Ingersoll_1956}& $3.46 \pm 0.04$ & Interpolated with\\
& & $A/\lambda^2+2B/\lambda^4$.\\
& & Scaled to water.\\
{This work}
& $3.56 \pm 0.10$ &  \\
 \hline
 \textbf{Theory} &  & \\
  & & \\
Savukov\cite{Savukov_2012} &  $3.86\pm 0.01$ & Interpolated in this work\\
& &with $A/\lambda^2+B/\lambda^4$.\\
Ekstr{\"o}m \textit{et al}\cite{ekstrom_2005} &   3.35 & TDHF\\

Ekstr{\"o}m \textit{et al}\cite{ekstrom_2005} &   3.46 & TDDHF\\

Ik\"al\"ainen \textit{et al}\cite{Ikalainen_2012}& 3.34 & NR\\

Ik\"al\"ainen \textit{et al}\cite{Ikalainen_2012}& 3.48 & X2C\\

Ik\"al\"ainen \textit{et al}\cite{Ikalainen_2012}& 3.46 & DHF\\
Ik\"al\"ainen \textit{et al}\cite{Ikalainen_2012}& 3.52 & NR-CCSD\\
{This work}    &  3.49 &  CCSD/t-aug\_cc\_pvqz\_pp\\
{This work}    &  3.52 &  CC3/t-aug\_cc\_pvqz\_pp \\
 \hline
 \hline
\end{tabular*}
\end{center}
\end{table}

The uncertainty of a few percent obtained on our experimental
value allows to comment on the agreement with theoretical
predictions as a function of the theoretical approximation or
model. %I.\,M.\,
Savukov\cite{Savukov_2012} has used a relativistic particle-hole
configuration interaction (CI) method. He does not give a value at
1064\,nm, but the latter can be interpolated, as done with the
previous experimental data of Ingersoll and
Liebenberg,\cite{Ingersoll_1956} obtaining the value of
Tab.\,\ref{Tab:ComparaisonFaraday}, with an uncertainty given by
the fit. The agreement between theory and experiment is only
within 3$\sigma$, even if relativistic effects are taken into
account. Ekstr\"om \textit{et al}\cite{ekstrom_2005} have used the
nonrelativistic time-dependent Hartree-Fock (TDHF in
Fig.\,\ref{Fig:Comparaison_Faraday}) and the relativistic
time-dependent Dirac-Hartree-Fock (TDDHF in
Fig.\,\ref{Fig:Comparaison_Faraday}). There is clearly a better
agreement (better than 1$\sigma$), between their calculations and
our experimental value when relativistic effects are taken into
account. Finally, Ik\"al\"ainen \textit{et
al}\cite{Ikalainen_2012} have used the non-relativistic
Hartree-Fock method (NR in Fig.\,\ref{Fig:Comparaison_Faraday}),
the exact two-component method (X2C in
Fig.\,\ref{Fig:Comparaison_Faraday}), and the fully relativistic
four-component method (DHF in
Fig.\,\ref{Fig:Comparaison_Faraday}). The same authors also report
(in the supporting information file) a non relativistic CCSD
result (NR-CCSD in Fig.\,\ref{Fig:Comparaison_Faraday}). While
their uncorrelated results confirm that relativistic effects
should be taken into account to improve agreement with experiment,
their non-relativistic CCSD result highlights how the inclusion of
correlation effects is equally important. Also worth noticing is
the rather poor performance of the BLYP and B3LYP functionals,
which overestimate the value of the Verdet constant in both
non-relativistic and relativistic calculations. This also applies
for the BHandHLYP functional in  the relativistic calculations,
whereas the non-relativistic BHandHLYP value is still within
1$\sigma$ of our experimental result (See Table S5 of the
Supporting Information file of Ref.~\citenum{Ikalainen_2012}).

\subsection{Cotton Mouton Effect}

\subsubsection{Experiments}

\begin{figure}
\begin{center}
\includegraphics[width=8cm]{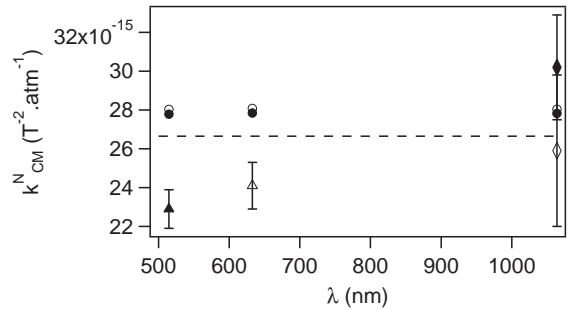}
\caption{\label{Fig:Comparaison_CM} Reported values of
Cotton-Mouton constant of xenon for $\lambda$ ranging from
514.5\,nm to 1064\,nm and with 1$\sigma$ uncertainty. Experimental
values: black triangle: Carusotto \textit{et
al},\cite{Carusotto_1984} open triangle: H\"uttner (private
communication by Bishop \textit{et al.}),\cite{Bishop_1991} black
diamond: Bregant \textit{et
al.}\cite{Bregant_2009_xenon,Bregant_2009_erratum}, open diamond:
this work. Theoretical predictions: dashed line: SCF method for
$\lambda = \infty$ by Bishop,\cite{Bishop_1993} open circle: this
work, CCSD, black circle: this work, CC3}
\end{center}
\end{figure}

Only a few measurements of the Cotton-Mouton effect in xenon have
been discussed in the literature. There is one at $\lambda =
514.5$\,nm by Carusotto \textit{et al},\cite{Carusotto_1984} one
at $\lambda = 632.8$\,nm by H\"uttner (reported as a private
communication by Bishop \textit{et al}),\cite{Bishop_1991} and
finally one at $\lambda = 1064$\,nm by Bregant \textit{et
al}.\cite{Bregant_2009_xenon,Bregant_2009_erratum} Our
experimental value, referring to $\lambda = 1064$\,nm is
compatible within 1$\sigma$ with the datum of Refs.
\citenum{Bregant_2009_xenon,Bregant_2009_erratum}. The set of
results is shown in Tab.\,\ref{Tab:ComparaisonCM} and plotted as a
function of the wavelength in Fig.\,\ref{Fig:Comparaison_CM}.

\begin{table}
\caption{\label{Tab:ComparaisonCM} Experimental (uncertainties of
1$\sigma$) and theoretical values of the Cotton-Mouton constant of
xenon at $T = 273.15$\,K.}
\begin{center}
%\begin{tabular*}{1\textwidth}{@{\extracolsep{\fill}}{c c c c c c}
\begin{tabular*}{0.47\textwidth}{p{4.5cm} p{1.5cm}  p{5cm}}
  \hline
  \hline
  Ref. & $\lambda$\,(nm)  & $k_\mathrm{CM}^{\mathrm{N}}\times10^{14}$\\
  & &(T$^{-2}$.atm$^{-1})$\\
 \hline
 \textbf{Experiment} & &  \\
 & & \\
 Carusotto \textit{et al} \cite{Carusotto_1984} & 514.5 & $(2.29 \pm 0.10)$\\
 %Bishop \textit{et al}\cite{Bishop_1991} & 632.8 & $(2.41 \pm 0.12) \times 10^{-14}$\\
 H{\"u}ttner\cite{Bishop_1991} & 632.8 & $(2.41 \pm 0.12)$\\
 Bregant \textit{et al}\cite{Bregant_2009_xenon,Bregant_2009_erratum} & 1064 & $(3.02 \pm 0.27)$\\
 This work & 1064 & $(2.59 \pm 0.40)$\\
 \hline
 \textbf{Theory} & & \\
 & & \\
 Bishop \textit{et al}\cite{Bishop_1993} & $\infty$ & $2.665$\\
 This work,  & 514.5 & $2.803$ \\
 CCSD/\mbox{t-aug\_cc\_pvqz\_pp}& &\\
 This work, & 632.8 & $2.808$ \\
 CCSD/\mbox{t-aug\_cc\_pvqz\_pp}& &\\
 This work, & 1064 & $2.804$ \\
 CCSD/\mbox{t-aug\_cc\_pvqz\_pp}& &\\
 This work, CC3/\mbox{t-aug\_cc\_pvqz\_pp} & 514.5 & $2.778$ \\
 This work, CC3/\mbox{t-aug\_cc\_pvqz\_pp} & 632.8 & $2.784$ \\
 This work, CC3/\mbox{t-aug\_cc\_pvqz\_pp} & 1064 & $2.782$ \\
 \hline
 \hline
\end{tabular*}
\end{center}
\end{table}

Our measurement has an uncertainty of about 15$\%$. This value,
which is larger than that of the other reported values, especially
those given for wavelengths of 514.5\,nm and 632.8\,nm, was
established via a complete error budget. Note that no information
is available on the setup, the number of pressures, the error
budget and the evaluation of the uncertainty for the value
reported at $\lambda = 632.8$\,nm by Bishop \textit{et
al}\cite{Bishop_1991} as a private communication of H\"uttner. The
value reported at $\lambda = 514.5$\,nm by Carusotto \textit{et
al}\cite{Carusotto_1984} was measured only at 1 atm, and by
comparing the observed magnetic birefringence with that of
nitrogen under the same experimental conditions, therefore taking
as a reference, free of uncertainty, the Cotton-Mouton constant of
nitrogen. It is safe to say therefore that the uncertainty
associated to their datum might be underestimated. Finally, the
value reported by Bregant \textit{et
al}\cite{Bregant_2009_xenon,Bregant_2009_erratum} at $\lambda =
1064$\,nm corresponds to the weighted average between measurements
at two different pressures (9 pressures for our measurement) and
the uncertainty is similar to ours.

\subsubsection{Theory}
The Cotton-Mouton constant $k_{\mathrm{CM}}$ is linked to $\Delta
\eta$ by the relationship\cite{Rizzo_Rizzo}
\begin{eqnarray}\label{Eq:delta_eta}
k_{\mathrm{CM}}\ (\mathrm{atm^{-1}T^{-2}}) = \frac{6.18381 \times
10^{-14}}{T} \Delta \eta\ (\mathrm{a.u.}).
\end{eqnarray}

Only one theoretical prediction has been published so-far for the
Cotton-Mouton effect in xenon.\cite{Bishop_1993} The calculation
of Bishop and Cybulski was performed at the self-consistent-field
(SCF) level of approximation, and it yielded static
hypermagnetizability anisotropy $\Delta \eta$. As stated by the
authors, relativistic effects were not taken into account, even
though the authors expected them to play a substantial role. Our
experimental value agrees with that theoretical prediction within
$1\sigma$.

Our computed coupled cluster results, both CCSD and CC3, in the
largest (t-aug\_cc\_pvqz\_pp) basis sets  for the three
wavelengths at which experimental results are given in
Tab.~\ref{Tab:ComparaisonCM}. Both the CCSD and CC3 values at 1064
nm are well within $1\sigma$ of our experimental measurement, and
just outside $1\sigma$ of the result by Bregant {\em et
al}.~\cite{Bregant_2009_xenon,Bregant_2009_erratum} At 632.8 nm
the agreement of our CC3 value with the experimental result of
H{\"u}ttner~\cite{Bishop_1991} is just outside $3\sigma$. At 514.5
nm our computed values fall well outside $3\sigma$ of the estimate
of Carusotto \textit{et al}.\cite{Carusotto_1984} This apparently
confirms that the error associated to this measured value might be
underestimated.

%%%%%%%%%%%%%%%%%%%%%%%%%%%%%%%%%%%%%%%%%%%%%%%%%
\section{Conclusion}
%%%%%%%%%%%%%%%%%%%%%%%%%%%%%%%%%%%%%%%%%%%%%%%%%

We have carried out a thorough analysis of the Faraday (circular)
and Cotton Mouton (linear) birefringences of xenon, at a
wavelength of 1064 nm. The study involves both an experimental
segment, exploiting the capabilities of a state-of-the-art optical
setup, and a computational element, where sophisticated
wavefunction structure and optical response models (and with an
estimate of the effect of relativity) were employed.

Our experimental estimate for the normalized Verdet constant of
xenon at a temperature of 273.15 K and $\lambda$=1064 nm,
$V^\mathrm{N}$ = (3.56$\pm$0.10) $\times 10^{-3}$ atm$^{-1}$ rad
T$^{-1}$ m$^{-1}$, is very well reproduced by our theoretical
approach, which yields a value ($V^\mathrm{N}$ = 3.52 $\times
10^{-3}$ atm$^{-1}$ rad T$^{-1}$ m$^{-1}$ using the CC3
approximation) within 1$\sigma$ of the measured datum.

With respect to the Cotton Mouton effect, at $T$=273.15 K and
$\lambda$=1064 nm, experiment yields a normalized constant
$k^\mathrm{N}_{\mathrm{CM}}$ = (2.59$\pm$0.40) $\times 10^{-14}$
atm$^{-1}$ T$^{-2}$, whereas we compute (again with our most
sophisticated model, CC3) a value of $k^\mathrm{N}_{\mathrm{CM}}$
= 2.78 $\times 10^{-14}$ atm$^{-1}$ T$^{-2}$, therefore within
1$\sigma$ of experiment.

%%%%%%%%%%%%%%%%%%%%%%%%%%%%%%%%%%%%%%%%%%%%%%%%%
\begin{acknowledgments}
%%%%%%%%%%%%%%%%%%%%%%%%%%%%%%%%%%%%%%%%%%%%%%%%%
We thank all the members of the BMV collaboration, and in
particular J. B\'eard, J. Billette, P. Frings, B. Griffe, J.
Mauchain, M. Nardone, J.-P. Nicolin and G. Rikken for strong
support. We are also indebted to the whole technical staff of
LNCMI. Sonia Coriani acknowledges useful discussions with Lucas
Visscher and Trond Saue. We acknowledge the support of the
\textit{Fondation pour la recherche IXCORE} and the \textit{Agence
National de la Recherche} (Grant No. ANR-14-CE32-0006).
\end{acknowledgments}

\bibliography{biblio}% Produces the bibliography via BibTeX.

\end{document}